\documentstyle[art11,epsf]{article}
\newcommand{\none}{\multicolumn{2}{c|}{-}}
\newcommand{\bnone}{\multicolumn{2}{|c|}{-}}
\def\prod{\mathop{\displaystyle \Pi}\limits}
\title{Instantons and Monopoles in the Maximally Abelian Gauge.}
\author{A. Hart\thanks{e-mail: {\sl harta@thphys.ox.ac.uk}}\, and M. Teper.
\\
{\small\sl Theoretical Physics, University of Oxford,
1 Keble Road, Oxford, OX1 3NP, U.K.}}
\date{November 10, 1995.}
\begin {document}
\maketitle
\begin{abstract}
\noindent
We study the Abelian projection of $SU(2)$ instantons in the
Maximally Abelian gauge. We find that in this gauge an
isolated instanton produces a closed monopole loop within
its core and the size of this loop increases with the core size.
We show that this result is robust against the introduction of small
quantum fluctuations. We investigate the effects of
neighbouring (anti)instantons upon each other and show 
how overlapping (anti)instantons can generate larger
monopole loops. We find, however, that in fields that are typical 
of the fully quantised vacuum only some of the large monopole 
loops that are important for confinement have a topological origin.
We comment on what this may imply for the role of instantons 
in confinement and chiral symmetry breaking.
\end{abstract}

\vspace{15ex}
Oxford Preprint Number: {\em OUTP--95--44--P}
\hfill
hep-lat/9511016
\vfill

\section{Introduction}

The idea of 't Hooft  
\cite{tHo}  
that confinement in non-Abelian gauge
theories might be associated with monopoles in suitable Abelian
projections of the fields, has been the subject of extensive
numerical investigation in recent years. This has been
largely due to the observation 
\cite{suzuki}
that in the Maximally Abelian gauge 
\cite{MAG}
the string tension one obtains from the Abelian Wilson loops
appears to equal the full non-Abelian string tension
\cite{magrev}.

Recently it has been shown analytically
\cite{CnG}
that if one puts a particular classical instanton potential into 
this gauge then one finds a monopole whose world line passes
through the centre of the instanton. This is interesting since 
large monopole loops contribute to (Abelian) confinement; but it
is puzzling because it is known that a reasonably decorrelated
ensemble of (anti)instantons does $not$ contribute to non-Abelian 
confinement. At the same time, because instantons probably contribute
to chiral symmetry breaking
\cite{diakonov}
\cite{hands},
this raises the possibility of learning something about the 
connection between chiral symmetry breaking and confinement.

The results of
\cite{CnG}
were obtained starting with an $SU(2)$ instanton in a particular
gauge and in an infinite volume. However it is well known that
the details of the monopole `gas' differ for the different Gribov 
copies of a given field. So the first question we would like to answer 
is whether the analytic result of 
\cite{CnG}
is typical of all these Gribov copies, or whether it is effectively 
of measure zero. Since the vacuum contains a `gas' of topological
charges we would also like to know how the presence of neighbouring 
charges affects the monopole properties of an instanton. Finally,
we obviously need to know how robust are our results against the
inclusion of quantum fluctuations about the instantons.

To answer these questions we shall discretise instantons on
hypercubic periodic lattices. We shall go to the Maximally
Abelian gauge numerically and will then study the monopole
content of the resulting Abelian fields. We add quantum
fluctuations to a lattice instanton by performing a
sequence of Monte Carlo steps with the instanton  as the initial
configuration. 

The purpose of our calculations is to tell us how instantons contribute
to the gas of monopoles, when one performs the Maximally Abelian 
projection of $SU(2)$ fields in which the instantons are 
the $only$ non-perturbative fluctuations. This is a well-defined
and tractable problem and its solution seems to us to be a 
prerequisite to answering the much more ambitious (although
probably ambiguous) question of how instantons contribute to 
the monopole gas in the fully quantised vacuum. To provide
the reader with an indication of what this latter problem entails 
we finish the paper by taking a few typical Monte Carlo generated field
configurations and calculate the corresponding monopole distributions. 
We then cool these fields, gradually exposing the topological charge
\cite{MTcool}
therein. We calculate the monopole distributions corresponding
to these smoothened fields and compare them to those of the
original fields and see if there are any correlations
with the topological charges in those fields. 

\section{Single Instantons}

To construct an instanton \cite{polyinst} on a discretised hypertorus
we follow the procedure in \cite{MTinst}. 
We begin with the gauge potential 
$$ A_{\mu} = { {x^2} \over {x^2 + \rho^2} }
g^{-1}(x) \partial_{\mu} g(x)  \eqno(1) 
$$
where
$$
g(x) = {{x_0 {\bf 1} + i x_j \sigma_j} \over \sqrt{x^2}},
\eqno(2)
$$
This corresponds to an instanton of size $\rho$ in a particular 
gauge and in an infinite volume. To construct a corresponding
lattice field we define the link variables by
$$
U_\mu(n) = {\cal P} {\hbox{exp}} \left( i \int_n^{n+\widehat{\mu}} 
dx_\nu A_\nu(x) \right) \eqno(3)
$$
(Note that this implies that $\rho$ is in units of the lattice spacing.)
In practice we calculate this by the product of $N$ discrete elements
$$
U_\mu(n) = \prod_{m=0}^{N-1} {\hbox{exp}}  \left( {i \over N}
A_\mu(n+{m \over N}\widehat{\mu}) \right)\eqno(4)
$$
where N is chosen sufficiently large for this to be an accurate
approximation. (For the particular solution we are using
eqn(3) can in fact be evaluated analytically.)
We now have an approximate lattice instanton
but it is in an infinite volume. Since
the fields go to different values as $x^2 \to \infty$
in different directions, we cannot simply take some
reasonably large volume centered on the instanton and impose
periodic boundary conditions on it. (If we do so then we
find that the mismatch between the fields at the boundary
typically carries a topological charge that cancels that of our
instanton --- so that the total field configuration in effect
possesses zero charge.) We therefore apply to the
lattice field the gauge transformation $g^{\dag}(n)$, and this
produces a potential that goes to $A_{\mu}(x)=0$ and hence
$U_{\mu}(n)=1$ at $x^2=\infty$. (In doing so we have gone
to a singular gauge. For numerical reasons it is
better to do so after discretising the instanton
field rather than before.) If we now `cut out' a volume
$L^4$ centered on the instanton and impose periodicity
on that volume, there will be only a small mismatch at
the boundaries as long as $L \gg \rho$. To smoothen
this approximate lattice instanton we typically perform
5 cooling sweeps upon it. As long as $L \gg \rho \gg 1$ 
the size of the instanton will not change significantly
and this final field is, to a good approximation,
a periodic lattice instanton of size $\rho$. (In practice
one finds that the cooling does not appreciably alter
the size of the inatanton as long as $\rho \geq 2$.) 

The Maximally Abelian (MA) gauge is defined as the gauge
in which the `adjoint' operator
$$ X(n) = \sum_{\mu} \{
U_{\mu}(n) \sigma_3 U^{\dag}_{\mu}(n)+    
U^{\dag}_{\mu}(n- \widehat{\mu} ) \sigma_3 U_{\mu}(n- \widehat{\mu} ) \}    
\eqno(5)
$$
points in the $\sigma_3$ direction for all $n$. Since this operator
depends on gauge transformations in a non-local way 
this can only be achieved by an iterative site-by-site method
which is made more efficient by incorporating over-relaxation
\cite{MnO}. 
Once $X(n)$ has been made proportional to $\sigma_3$
there is a remnant $U(1)$ symmetry which leaves this
invariant and this defines for us our $U(1)$ fields. These
fields will generically contain topological singularities that are
magnetic monopoles. The corresponding
magnetic 4-currents on the dual Abelian lattice can be identified 
\cite{DeGT}
and resolved into separate closed monopole world-lines. (This last
step is largely unambiguous in our case, since typically only 1-2\%
of the dual links carry a non-zero monopole current.)

We have constructed lattice instantons of various sizes 
and on various lattice volumes using the above construction.
When we put these fields into the MA gauge, we find that  
each of the corresponding U(1) fields contains a single monopole 
loop (as long as the instantons are not too small: $\rho \geq 2$). 
This loop lies within the core of the instanton and its 
length is an increasing function of the instanton core size,
although to see this one must be on lattices that are
sufficiently large; see Table \ref{onelens}. Moreover the loop 
tends to be planar on the smaller lattices and for
the smaller values of $\rho$. As one would expect, the 
centroid of the loop (nearly) coincides with the 
centre of the instanton as measured by the maximum of $Q(n)$. 

Of course the above results have been obtained for the initial
instanton configuration in a very special gauge. As we
have previously remarked, the extracted $U(1)$ fields
differ between differing Gribov copies and so it is possible
that what we have seen so far is not characteristic of general
instanton fields. To investigate this question
we generate more general instanton fields by applying 
random $SU(2)$ local gauge transformations to the
lattice instantons of eqns(1-4). When we do so
we once again find the characteristic feature
that each such field possesses a single monopole loop
(if the instanton is not too small). There are, however, 
variations in the lengths and shapes of the loops and
these variations become larger for larger lattices
and larger instantons. In Table \ref{oneglens} we
present the average loop lengths and the standard
deviations as obtained from 5 such randomly
gauged instanton fields for each instanton and lattice
size. We can infer that for a fixed instanton size the
loop length becomes independent of the lattice size, $L$,
once $L \geq 2\rho$. We also see that this loop length
grows with $\rho$ and that this growth is roughly linear.
This is not unexpected given the scale-invariance of
the continuum gauge theory in an infinite volume.

At this stage we have established what occurs when we
have a single classical instanton. If this is to be
of relevance to the quantum field theory, we must show
that what we have seen so far is in fact robust 
under the inclusion of, at the very least, small quantum 
fluctuations. To address this question we need to 
incorporate such fluctuations around the instanton field.
To do so we start with our usual classical lattice 
instanton field and then perform a few Monte Carlo sweeps 
at some value of $\beta$. The number of sweeps needs
to be chosen large enough that perturbative 
fluctuations are accurately incorporated (in the sense 
that the plaquette is close to its equilibrium value) 
while being small enough that the background instanton field 
is essentially unchanged (as can be checked by cooling
the final configuration). The value of $\beta$ is
chosen so that the lattice volume being used
is sufficiently small that we do not run the risk
of generating other large-scale non-perturbative fluctuations.
The results we present here have actually been obtained
using 5 Monte Carlo sweeps at $\beta=3.0$ on a
$16^4$ lattice. 

\begin{figure}
\begin{center}
\leavevmode
\epsfxsize=6in
\epsffile{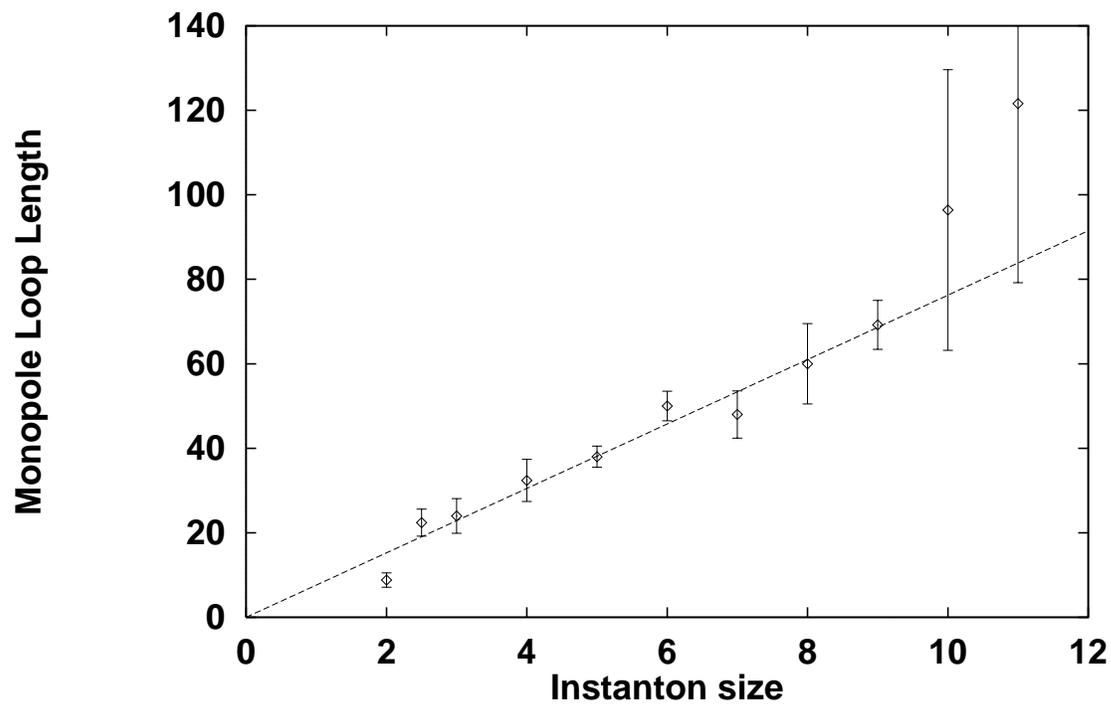}
\end{center}
\caption{Monopole loop length versus instanton size for instantons,
including quantum fluctuations, on a $16^4$ lattice.}
\label{plot}
\end{figure}

We once again find that an instanton is always associated
with a monopole loop located within the core of the
instanton. We show in Fig. \ref{plot} how the average loop
length varies with instanton size. The errors are based on
5 different field configurations at each instanton size
and so should only be taken as indicative of the true errors.
Nonetheless we find clear evidence for a linear
dependence between instanton size and loop length, even
in the presence of quantum fluctuations which ultimately
break the scale-invariance of the theory. The loop length
is typically much longer than for the corresponding classical
case, as we see if we compare Table \ref{oneglens} and 
Fig \ref{plot}, and this seems to be primarily because 
the perimeter of the loop is much less smooth.
In addition to the loop associated with the monopole, there
are typically several much smaller loops in the vacuum - and
this is so even if we have no instanton. Such small loops
will, however, play no role in affecting phenomena on
physical length scales. If the field is now cooled then
the loop shrinks back close to its characteristic  classical
length.

By now it should be clear that our results differ 
from those of \cite{CnG}. This is no surprise since 
that solution $had$ to be special to a strictly infinite volume.
The reason is that any 3-volume orthogonal to the monopole world 
line found in \cite{CnG} contains just a single static monopole 
and, by the conservation of magnetic flux, this would
not be possible on a finite periodic volume. A contractible
monopole loop, in contrast, will correspond to a 
monopole-antimonopole pair in any intersecting 3-volume.

\section{Interacting (Anti-)Instantons}

The real vacuum will contain a number of topological charges that
is proportional to the volume. If the separations between these
charges were large compared to the typical core size, then
the calculations in the previous section would imply
that the corresponding $U(1)$ fields would consist of 
a dilute gas of monopole loops centered on these charges,
with sizes comparable to the sizes of the corresponding
cores. However in the real world the typical core size 
and typical inter-core
separation must be comparable because the theory contains
only one scale. In this case what happens to the monopole
loops? In particular do they form much larger loops and
so affect the confining properties of the theory?

To approach this question we first consider the simpler case
of an instanton and an (anti)instanton, each of size $\rho$
and separated by a distance $\delta$. Such a configuration
can be created by joining two hypercubic sub-lattices on which
instanton potentials have been separately constructed, and
performing say five cooling sweeps to smoothen any mismatch at
the boundaries. This method fails if $\delta$ is too small
as compared to $\rho$ ---  one finds that the end result is only
a single instanton (as one might expect). Similarly for
an instanton and anti-instanton.

We begin with topological charges constructed to be oriented along the
4-direction in group space, as in eqns(1,2),
and on sub-lattices arranged along the 4-direction of space.
After projecting to the MA gauge, we find that if $\delta$ is small
we do not have 2 separate monopole loops but rather a single 
larger loop that encloses both centres. The length is too
great to be accounted for as a simple superposition of two
separate loops. The separation below which 
this occurs is substantially less for pairs of like
charge than for ones of opposite charge 
[see Table~\ref{twolens}]. For like pairs the critical
separation at which the two individual monopole loops merge
is about equal to the core size while for unlike pairs
it is about twice as large. Different Gribov copies (obtained
by performing random gauge transformations on the fields
just before MA gauge fixing) do not differ in any essentials:
a large loop remains large although its plane may rotate
around the axis joining the centres of the instanton and
(anti-)instanton.

\begin{figure}
\begin{center}
\leavevmode

\hbox{
\epsfxsize=2.5in
\epsffile{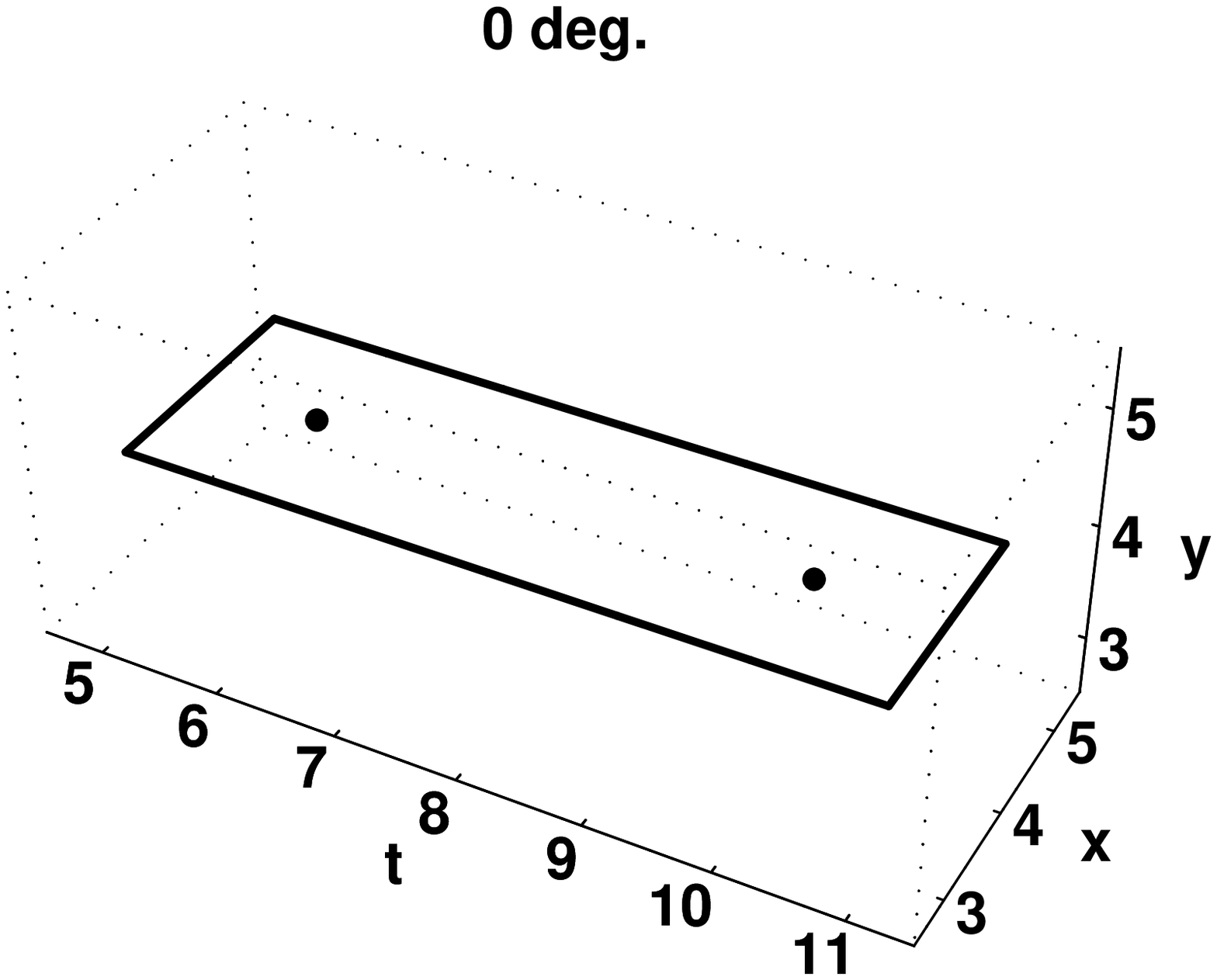}
\hspace{1in}
\epsfxsize=2.5in
\epsffile{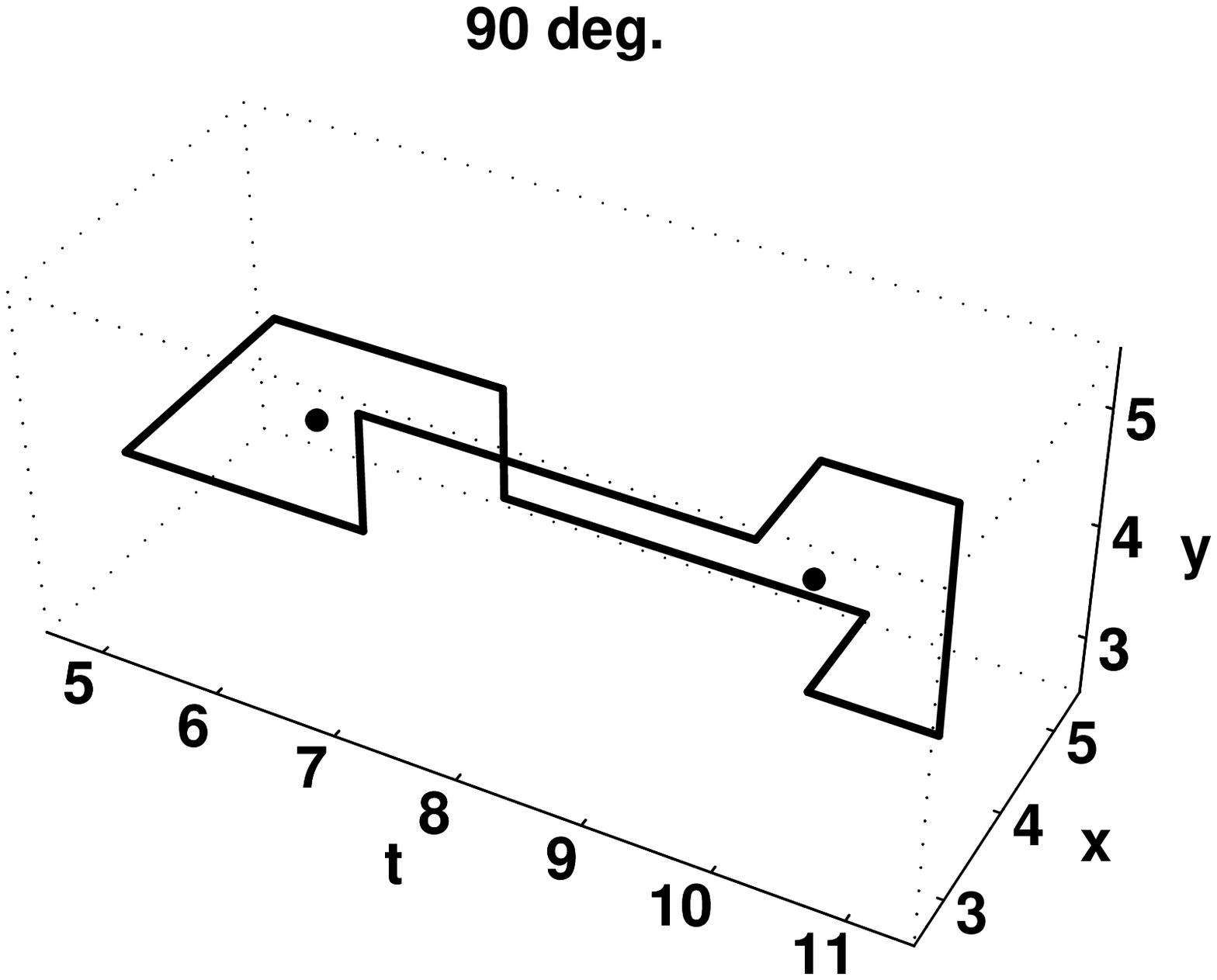}
}

\vspace{-1.3in}
\hbox{
\epsfxsize=2.5in
\epsffile{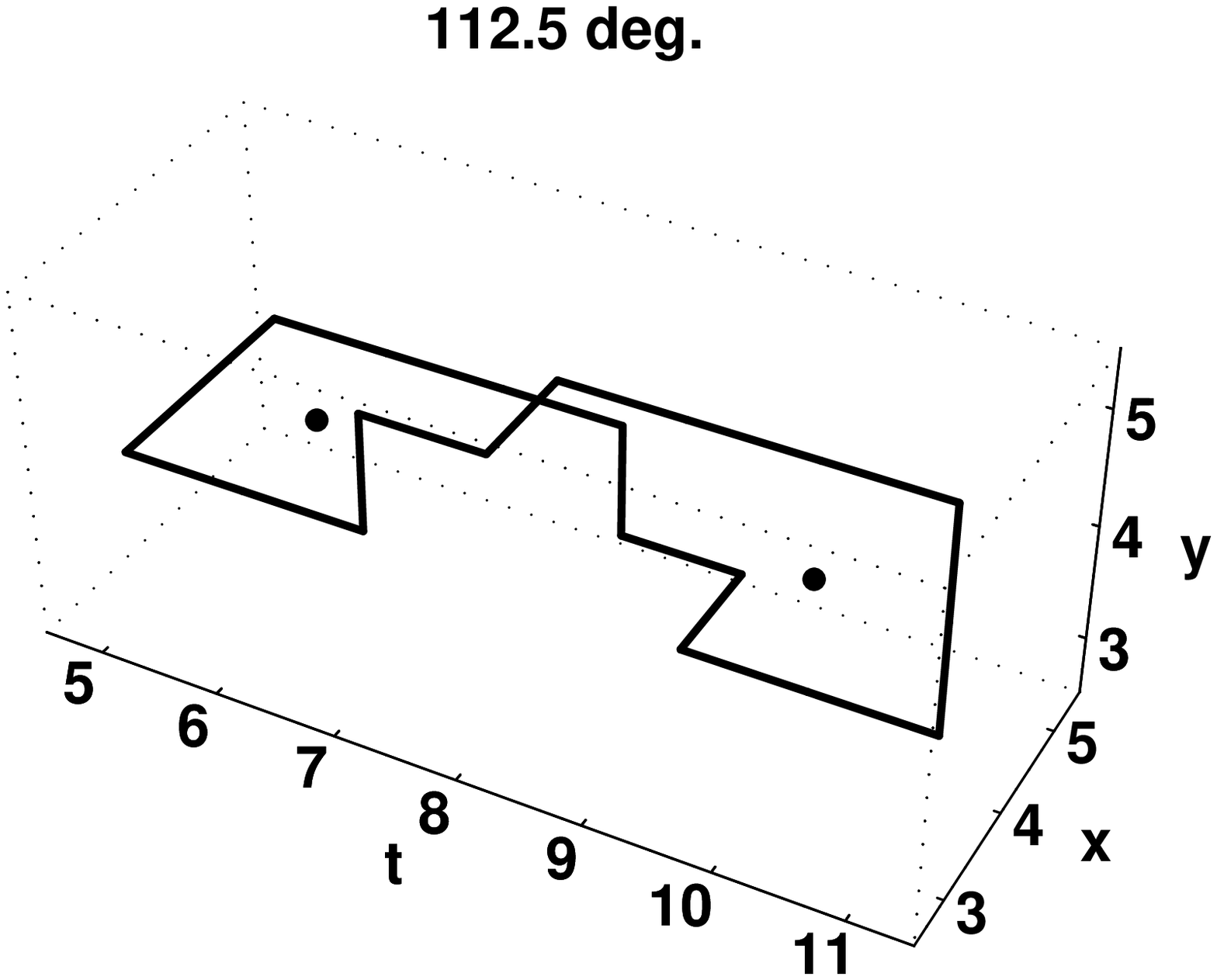}
\hspace{1in}
\epsfxsize=2.5in
\epsffile{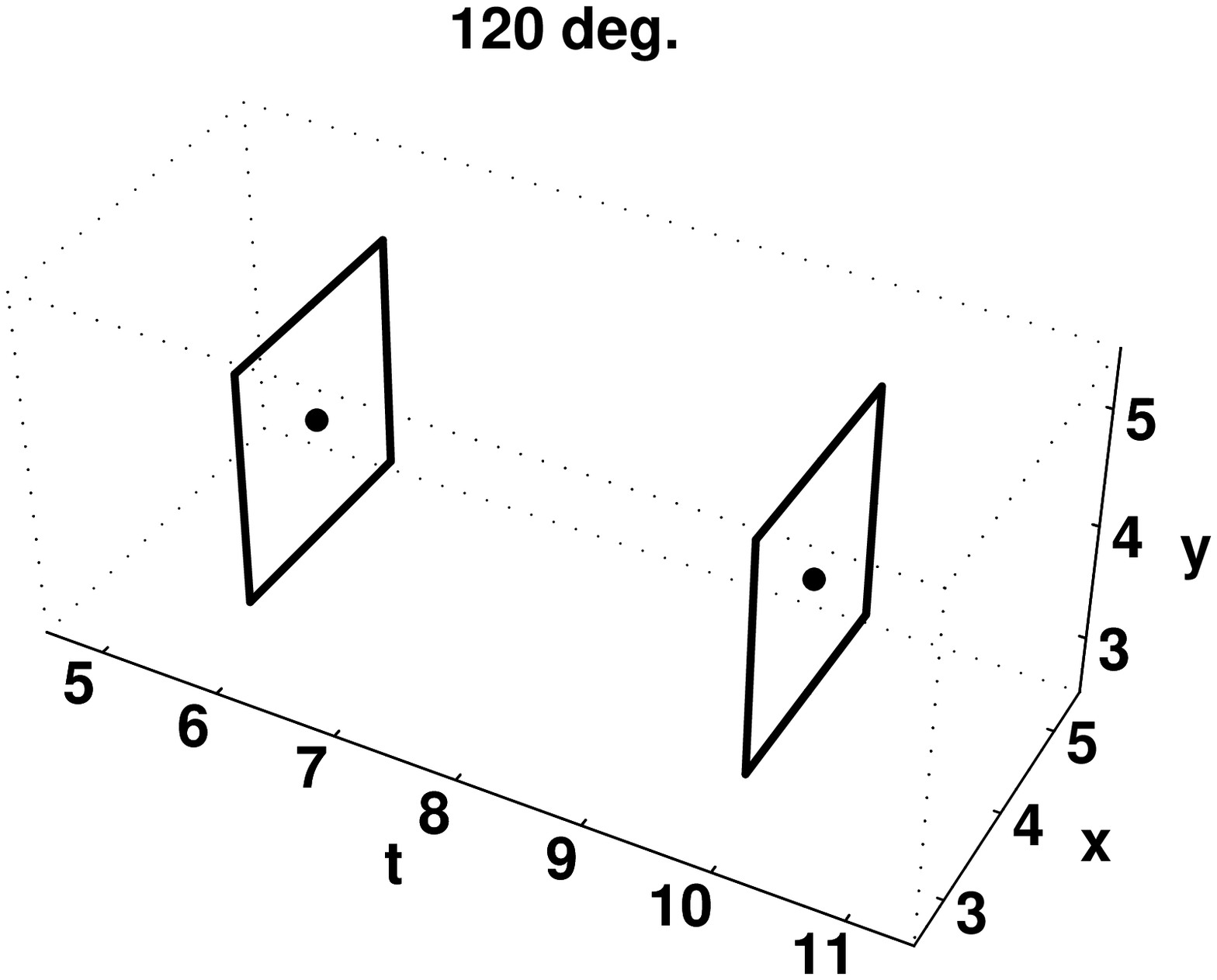}
}
\vspace{-1in}
\end{center}
\caption{Three dimensional projections of the mutual monopole loop
surrounding an instanton--anti-instanton pair (centres marked) of size
$\rho=3$ under increasing rotation angle as detailed in the text. The
loops are flat in the fourth direction.}
\label{twisting}
\end{figure}

In the real vacuum the relative orientations in real and group spaces
of neighbouring topological charges will not be the same, although
they may well be correlated, partly through their interactions (the
action will depend on the relative orientation) and partly through the
fact that they will be embedded within other nonperturbative
fluctuations. We have therefore investigated what happens under
changes in both the orientation in group space and relative
positioning of, say, an anti-instanton relative to that of the
neighbouring instanton. We have studied the five different relative
positionings of the sub-lattices along the principal lattice axes,
with rotations of the anti-instanton solution relative to that of the
instanton also about these axes. We focus on the interesting case
where $\delta$ is small enough to produce a single large monopole loop
for at least some orientations.  For the case described in the
previous paragraph,we find that for rotations up to about $\pi/6$ the
loop does not change. For larger angles it begins to twist near the
centre. It is only once the angle becomes greater than about $2\pi/3$,
however, that the monopole loop breaks into two separate loops, each
centered on one of the two topological charges [see
Figure~\ref{twisting}].  Similar angle dependent loop structure is
seen in two more of the five cases, which makes it reasonable to infer
that for completely general positioning and rotation, which we have
not investigated, nearby topological charges have a substantial
probability to form a single, large loop rather than two smaller ones.

We have carried out a less complete investigation for cases of more
instantons.
By creating a linear stack of three sub-lattices containing alternate 
instanton and anti-instanton solutions with equal separations of
centres, we find that for rotations of each sublattice by different
angles (although about parallel axes) there are orientations
for which a large single loop enclosing all three centres is formed 
rather than three single loops. For all five relative positions and
rotations the formation of this appears to be 
governed by the relative orientations of nearest neighbour charges 
only, and depends upon this in accordance with the pair-wise interactions
discussed above. Similar results are seen for linear stacks of four 
alternating instanton charges. By placing 4 sub-lattices together 
we obtain a $2 \times 2$ chequer--board of instantons and anti-instantons 
and we then find cases of two mutual loops each
enclosing a separate pair of neighbouring, opposite charges, or of one 
large loop enclosing all four. The latter case occurs when all of the four
possible instanton--anti-instanton pairs have relative orientations
consistent with formation of a mutual loop in the pair-wise
interaction. The former when this is so for only two such pairs.
A few trial cases of $2 \times 2 \times
2$ and $2 \times 2 \times 2 \times 2$ chequer--boards of opposite
charges yielded loops that could similarly be explained in terms of
pair-wise interactions of neighbouring charges although in these
cases we have not yet seen cases where there is only a single
large monopole loop. 

The above calculations suggest that if 
the instanton gas is moderately dense, as one would expect 
for a theory with one scale, then the monopole loops that would have
been associated with the individual topological charges might
merge into a smaller number of larger loops. This
would seem to require a rather specific ordering of neighbouring
instanton orientations. Such large loops, if formed, would
be interesting because they could have an impact on the confining
properties of the theory. Clearly a systematic study of
the multi-instanton case, including quantum fluctuations, would
be worth carrying out and we intend to do so.

\section{The $SU(2)$ Vacuum}

So far we have investigated the connection between monopoles
and instantons in fields where the former are the $only$
non-perturbative fluctuations. Eventually, however, we
would like to understand what happens in the fully quantised
vacuum. We therefore include here a brief and far from
systematic study of the real gauge vacuum.

To do so we take a typical $SU(2)$ field configuration
on a $16^4$ lattice generated at $\beta=2.5$. 
(Such a lattice is moderately large in physical units.)
We put this field into the MA gauge and extract the
corresponding monopole loops. We then perform a sequence
of cooling sweeps on the $SU(2)$ field. This rapidly
removes the high frequency fluctuations and exposes
the topological structure. (If the lattice were
arbitrarily large and the lattice spacing arbitrarily
small then the cooling would eventually produce
a multi-instanton configuration corresponding to
the minimum of the action.) As we cool, we put the
cooled configuration into the MA gauge and extract
the corresponding monopole loops. In Table~\ref{cooling} 
we show how the number of loops and
their average length varies with the number of $SU(2)$ cooling sweeps.
After the second cooling sweep there are only 3 loops. Two
are relatively small and they lie within the two
pronounced (because rather narrow) topological charges
in the cooled field. The third loop is very long 
(initially about 220 links), extends over the whole lattice
and shrinks only gradually with further cooling. There are
a number of broader, overlapping instantons and anti-instantons 
on the lattice at this stage of cooling and since there are no 
individual monopole loops associated with them, it 
must be that these loops have joined into a
much larger loop (as in the previous section) and
that this is just our third  loop. If we now continue cooling
then we find that somewhere between
10 and 15 cooling sweeps one of the narrowish instantons shrinks
out of the lattice and the corresponding small monopole loop
disappears. Eventually the other narrow instanton shrinks
out of the lattice and only one loop remains. By the time
we have performed 150 cooling sweeps our field contains only one
instanton and what was our very long third loop is the
only loop present, being now 34 links long and within the
core of the remaining instanton.

This kind of study (we have carried out additional examples) 
demonstrates in a direct way that while the smallest 
instantons will be associated with small monopole loops,
the monopole loops associated with the 
overlapping topological charges of a more typical
size do in fact  combine into very large loops that extend
over much of the lattice. Of course it may be that
they have combined to form only part of the large loop
in our example: it might be that there are other 
non-perturbative fluctuations that also contribute to the 
formation of this loop. One might object that
after a few cooling sweeps this would be an
unnecessary qualification. However this is not so.
To test this hypothesis we have cooled $SU(2)$ fields
in 2+1 dimensions where there are no non-trivial
minima of the action and where under the cooling the lattice 
action smoothly descends to zero. In that case we
find it common for cooled fields to possess distant
monopole-antimonopole pairs even after 10 cooling sweeps.
This tells us that we cannot assume that on the
cooled lattices any monopole loops must be of
topological origin.

Our ultimate interest is, of course, in the monopole loops 
associated with the original uncooled lattice field.
We have focussed on the cooled lattices 
because from these we can learn what kind of monopole loops
are associated with the topological structure of the fields.
Having done so, we can ask about the nature of those
loops that are not topological in origin. Our particular
interest is with the very extended monopole loops that
will disorder large Wilson loops and hence contribute
to confinement. Now, as we see
from the first column in Table~\ref{cooling}, 
the original configuration
contained a large number of loops. Most of these 
immediately disappeared under cooling. In addition
this included some very long loops. Of course it is possible
that some loops are long not because they are
very extended but because they are very crumpled and in
that case they would not contribute to confinement.
To address this question in a simple manner we
introduce a method to smoothen the $U(1)$ fields.
The simplest technique is to perform $U(1)$ cooling in a
manner entirely analogous to the $SU(2)$ cooling.
We sweep through the $U(1)$ lattice field minimising,
say, the plaquette action. This rapidly removes
the high frequency fluctuations of the $U(1)$
fields. Any kinks and crumpling in the monopole loops
is removed leaving large smooth monopole loops
that are nearly stable under further $U(1)$
cooling. In Table~\ref{cooling} we show
the number of monopole loops and their average length
as a function of the number of $U(1)$ cooling
sweeps performed on the $U(1)$ field obtained
from the original fully quantised $SU(2)$ field.
As we see from the Table most of the original loops must
be either small or crumpled because they disappear
almost immediately under the $U(1)$ cooling.
There are, however, several very large loops remaining,
in contrast to the single large loop associated with
the topological structure, and so it is clear that most
of the very large monopole loops that can be important for
confinement, do not in fact have a topological origin.
This conclusion has been reinforced by the study
of further cases than the one described in detail here.

\section{Conclusions}

We have shown that for cases where the only non-perturbative 
fluctuation in the $SU(2)$ fields is an isolated 
instanton, the $U(1)$ field obtained
by going to the Maximally Abelian gauge contains
a single monopole loop within the core of the instanton.
The size of the loop is proportional to the core size.
We have also studied instanton (anti)instanton pairs
as a function of their separation. For large separations
these simply produce two individual loops and from this we can
infer that a dilute gas of (anti)instantons will produce
a correspondingly dilute gas of finite monopole loops.
Such a gas of loops would not affect the confining properties 
of the theory, since that depends on the presence of
a suitable distribution of arbitrarily large loops.
On the other hand we find that for reasonably small separations, 
the loops typically do combine into larger loops
and this raises the possibility that a moderately dense
gas of topological charges, of the kind that one would
expect to be present in the real vacuum, could produce
very extended loops. Our study of examples of typical
vacuum fields suggested that while this indeed seemed to be the
case, only a fraction of the very large loops that
drive Abelian confinement, have a topological
origin. It is also the case that our calculations suggest
that the formation of such 
large loops from the background topological structure
requires that the correlation length characterising the
relative orientations (in group space) of nearby instantons
should not be too small. If these correlations are
produced by vacuum fluctuations other than instantons, as
they probably have to be, then
the association of the resulting very large loops with
the topological structure becomes less causal.

We finish with some speculative remarks on the
implications of this for the physics of the vacuum. Firstly,
if, as we have seen, the topological structure of 
the $SU(2)$ vacuum typically produces
monopole loops that extend over the whole space-time
volume, then it contributes to the linear confining
potential. This contradicts what is found with
direct calculations in instanton gas and liquid models. 
The solution to this puzzle probably lies in our
observation that the formation of large loops
requires strong correlations between the orientations, in
group space, of neighbouring charges. This would not
occur in an instanton gas or liquid and probably
requires the presence of other non-perturbative
fluctuations.

Our second speculation concerns  chiral symmetry breaking.
A fermion in the field of a static monopole has a
zero energy mode \cite{ttwu}. Thus one expects that a large
monopole loop should be associated with a small eigenvalue
of the Dirac operator, with this eigenvalue going to zero
as the size of the loop grows to infinity. So large loops
that extend throughout the lattice could be associated
with a distribution of modes that extends to zero. 
On the other hand in current lattice calculations 
one finds that it is the topological fluctuations
of the $SU(2)$ fields that produce the small modes that 
spontaneously break chiral symmetry \cite{hands}. 
So it would be natural
if in the $U(1)$ fields it was the large monopole loops 
associated with topology that provided the symmetry breaking
spectrum of near-zero modes. This is an idea that could
be tested explicitly. It certainly fits in with recent work 
\cite{wolo} showing that the $U(1)$ fields do possess a
symmetry breaking eigenvalue spectrum of the Dirac operator.

\section{Acknowledgements}

An earlier version of this work was presented at
the Workshop on Non-Perturbative QCD at ECT (Trento)
this summer and we are grateful for very useful discussions
with participants there. One of us (AGH) acknowledges 
the support of PPARC studentship 93300888. The calculations
were supported by PPARC Grants GR/J21408 and GR/K55752.

\newpage

\begin{table}
\begin{center}

	\leavevmode
	\hbox{

	\begin{tabular}[t]{|c||c|c|c|c|}
		\hline
		one $i $ & \multicolumn{4}{|c|}{Lattice size} \\
		\cline{2-5} 
		$\rho$       & $8^4$ & $12^4$ & $16^4$ & $24^4$ \\
		\hline\hline
		 2.0  		& 8	& 4	& 0	& -	\\
		 2.5  		& 8	& 8	& 8	& -	\\
		 3.0  		& 12	& 8	& 8	& -	\\
		 3.5  		& 16	& 8	& -	& -	\\
		 4.0  		& 8	& 8	& 8	& 16 	\\
		 5.0  		& 16	& 16	& 8	& -	\\
		 6.0  		& 10	& 16	& 16	& 16  	\\
		 7.0  		& 8	& 16	& 16	& -	\\
		 8.0  		& 8	& 16	& 16	& 24	\\
		 9.0  		& 8	& 16	& -	& -	\\
		 10.0  		& 8	& 16	& 16	& 24	\\
		 12.0  		& -	& -	& -	& 24	\\
		 14.0  		& -	& -	& -	& 32	\\
		\hline
	\end{tabular}
}
\end{center}
\caption{Lengths of monopole loops for various instanton and lattice
sizes, using the particular solution of equations (1,2) ( - denotes
this case unstudied).}
\label{onelens}
\end{table}

\begin{table}
\begin{center}
	\leavevmode
	\hbox{

	\begin{tabular}[t]{|c||r@{ }l|r@{ }l|r@{ }l|r@{ }l|r@{ }l|}
		\hline
		one $i$ & \multicolumn{10}{|c|}{Lattice size} \\
		\cline{2-11} 
		$\rho$ & \multicolumn{2}{|c|}{$6^4$} & 
		\multicolumn{2}{c|}{$8^4$} & \multicolumn{2}{c|}{$12^4$} &
		\multicolumn{2}{c|}{$16^4$} & \multicolumn{2}{c|}{$24^4$} \\
		\hline\hline
2.0 & 8.0&(0)	& 8.0&(0)	& 2.4&(2.2)	& 0&(0)		& \none \\
2.5 & 8.4&(0.9)	& 8.0&(0)	& 8.4&(0.9)	& 8.0&(0)	& \none \\
3.0 & 9.6&(0.9)	& 11.6&(0.9)	& 8.8&(1.1)	& 8.8&(1.8)	& \none \\
4.0 & 8.0&(0)	& 14.4&(2.2)	& 12.0&(2.5)	& 10.8&(1.1) 	& 10.4&(2.2)\\
5.0 & 8.0&(0)	& 13.2&(2.3)	& 15.2&(1.1)	& 16.0&(0)	& \none \\
6.0 & 8.0&(0)	& 11.2&(2.1)	& 16.0&(1.1)	& 15.2&(1.8)  	& 18.0&(7.9)\\
7.0 & \bnone	& 9.6&(1.7)	& 15.2&(1.8)	& 18.4&(0.9)	& \none \\
8.0 & \bnone	& 8.0&(0)	& 18.4&(0.9)	& 19.6&(0.9)	& 18.4&(3.6)\\
9.0 & \bnone	& \none		& 19.6&(0.9)	& 21.2&(2.7)	& \none \\
10.0& \bnone	& \none		& 21.2&(0.7)	& 22.4&(0.9)	& 34.8&(17.2)\\
11.0& \bnone	& \none		& 22.4&(0.9)	& 22.4&(2.6)	& \none \\
12.0& \bnone 	& \none		& 22.4&(2.6)	& \none		& 51.2&(24.8)\\
14.0& \bnone 	& \none		& \none		& \none		& 51.6&(11.9)\\
		\hline
	\end{tabular}
}
\end{center}
\caption{Lengths of monopole loops associated with instantons
of size $\rho$ in various lattice volumes: means (standard deviations).}
\label{oneglens}
\end{table}

\begin{table}
\begin{center}

	\leavevmode
	\hbox{
	\begin{tabular}[t]{|c||c|c|}
		\hline
		$i - i$ & \multicolumn{2}{|c|}{Lattice size} \\
		\cline{2-3} 
		$\delta$       & $8^3.16$ & $12^3.24$ \\
		\hline\hline
		12.0	& -     &  8,8  \\
		8.0     & -     &  8,8  \\
		7.0     & -     & -     \\
		6.0     &  8,8  & -     \\
		5.0     &  8,8  &  8,8  \\
		4.0     &  8,8  &  8,8  \\
		3.0     & 12    & 12    \\
		2.0	& -	& 12	\\
		\hline
	\end{tabular}

	\hspace{0.5in}

	\begin{tabular}[t]{|c||c|c|}
		\hline
		$ i - \bar {\imath}$ & \multicolumn{2}{|c|}{Lattice size} \\
		\cline{2-3} 
		$\delta$       & $8^3.16$ & $12^3.24$ \\
		\hline\hline
		12.0	& -     & 8,8   \\
		8.0     & 8,8   & -     \\
		7.0     & 8,8   & 8,8   \\
		6.0     & 8,8   & 24    \\
		5.0     & 18    & 18    \\
		4.0     & 16    & 20    \\
		3.0     & 10    & 10    \\
		2.0	& 0	& 0	\\
		\hline
	\end{tabular}
}
\end{center}
\caption{Lengths of monopole loops for (a) like and 
(b) unlike pairs of charges with $\rho = 3$.}
\label{twolens}
\end{table}

\begin{table}
\begin{center}
	\begin{tabular}{|c||c|c|c|c|c|c|c|c|c|c|}
	\hline
	$SU(2)$ Cools & 0 & 1 & 2 & 3 & 4 & 5 & 10 & 15 & 20 & 150 \\
	\hline \hline
	No. loops & 366 & 10 & 3 & 3 & 3 & 3 & 3 & 2 & 2 & 1 \\
	\hline
	Mean length & 9.7 & 42.0 & 82.7 & 78.0 & 75.3 & 74.7 & 64.0 & 
	85.0 & 81.0 & 34 \\
	\hline
	\end{tabular}

\vspace{1ex}
	\begin{tabular}{|c||c|c|c|c|c|c|c|c|c|c|}
	\hline
	$U(1)$ Cools & 0 & 1 & 2 & 3 & 4 & 5 & 10 & 15 & 20 & 150 \\
	\hline \hline
	No. loops & 366 & 23 & 17 & 16 & 14 & 11 & 9 & 8 & 8 & - \\
	\hline
	Mean length & 9.7 & 50.7 & 56.7 & 54.6 & 56.7 & 67.1 & 67.6 & 
	68.3 & 62.3 & - \\
	\hline
	\end{tabular}
\end{center}
\caption{Number of loops and average loop length of one typical 
configuration under cooling.}
\label{cooling}
\end{table}

\end{document}